# Tracking single *C. elegans* using a USB microscope on a motorized stage

Running Head: Single worm tracking

Eviatar I. Yemini[1a] and André E.X. Brown[2b]

[1]Howard Hughes Medical Institute, Department of Biological Sciences, Columbia University, USA
[2]MRC Clinical Sciences Centre, Faculty of Medicine, Imperial College London, UK

[a]eiy1@columbia.edu
[b]andre.brown@csc.mrc.ac.uk

## Summary

Locomotion and gross morphology have been important phenotypes for *C. elegans* genetics since the inception of the field and remain relevant.  In parallel with developments in genome sequencing and editing, phenotyping methods have become more automated and quantitative, making it possible to detect subtle differences between mutants and wild-type animals.  In this chapter, we describe how to calibrate a single worm tracker consisting of a USB microscope mounted on a motorized stage and how to record and analyze movies of worms crawling on food.  The resulting quantitative phenotypic fingerprint can sensitively identify differences between mutant and wild type worms.

## 1. Introduction

Digital cameras are becoming smaller and cheaper, a trend that is driven in part by the mobile phone industry and that is likely to continue for some time.  This trend can be exploited for worm tracking using mass-market USB microscopes as the core element of an automated motorized microscope.  There are now many published worm trackers designed for a variety of applications with associated advantages and challenges.  For a review of current worm trackers, see Housson *et al.* (1).

In this chapter, we will describe how to use a recently published single-worm tracker (2), including the hardware setup, basic experimental protocol, and data analysis.  The main advantages of this tracking system are that it provides a high-resolution view of a single worm over time and that the subsequent analysis is automated and unbiased.  Because the system is relatively inexpensive, we have been able to operate eight single worm trackers in parallel to increase throughput.

Automated analysis of experimental data often requires a compromise between the effort put into optimizing experimental conditions and the effort put into algorithm design: clean data can be difficult to collect but easy to analyze and *vice versa*.  In the case of single worm tracking, we have found that routine calibration and care during plate preparation is worth the small extra effort and makes the analysis relatively straightforward and robust. The most important factors for successful single worm tracking are reproducible sample preparation, a thin lawn of bacterial food, and uniform lighting adjusted to give good contrast.

The expected outcome of following the method described in this chapter is a quantitative phenotype with both morphological and behavioral features that can be used to distinguish worm strains, even when the differences between them are too subtle to observe by eye.

## 2. Materials

### *2.1 Worm Tracker Hardware*

The single worm tracker used in this protocol is called Worm Tracker 2 (WT2). The hardware consists of a worm platform, USB microscope, motorized stage, and red-light illumination. Single worms are placed into agar petri dishes. The dishes are placed on an immobile platform, shielding the worm from stage movements. The USB microscope is coupled to ~627nm LED illumination so as to permit imaging without triggering worm avoidance behaviors (*C. elegans* sense and avoid short-wavelength light (3)). Both the camera and its illumination are mounted onto a motorized stage. A closed video loop permits the tracking software to guide the stage in pursuit of the worm while recording video of its behavior.

1. DinoLite AM413T USB microscope
2. Motorized stage (e.g. Zaber TSB60-M translation stage plus corresponding linear actuators)
3. A red light-emitting diode (620-645nm wavelength)
4. A computer to control the stage and camera

Figure 1 shows an assembled worm tracker.

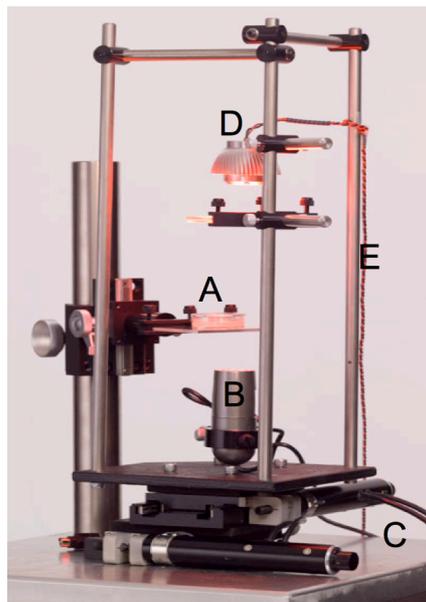

Fig. 1. WT2 Tracking Hardware. (**A**) A worm on a plate is kept still on a separate platform. (**B**) The DinoLite USB microscope is mounted directly onto the stage. (**C**) A

Zaber motorized stage is used to direct the camera, in order to follow the worm. (**D**) A red LED is mounted above the plate for illumination. (**E**) An opal diffuser diffuses the LED to create uniform lighting within the camera's field of view.

### *2.2 Worm Tracker Software*

In addition to a computer running Windows, the worm tracker uses the following software:

1. DinoCapture, included with the DinoLite USB microscope
2. Java software version 6 or later (http://www.java.com)
3. Worm Tracker 2 and Worm Analysis Toolbox, available at http://www.mrc-lmb.cam.ac.uk/wormtracker/

### *2.3 Agar/Nematode Growth Medium Plates*
(this subsection is adapted from WormBook.org (4))

1. In a 2l Erlenmeyer flask, add 3g NaCl, 17g agar, 2.5g peptone, and 975ml of distilled water. Autoclave for 50 min.
2. Cool flask in a water bath at 55$^o$C for 15 minutes.
3. Add 1ml of 1M $CaCl_2$, 1ml of 5mg/ml cholesterol in ethanol, 1ml of 1M $MgSO_4$, and 25ml of 1M $KPO_4$ buffer (108.3g of $KH_2PO_4$, 35.6g $K_2HPO_4$, made up to 1l in distilled water). Swirl to mix. This is Nematode Growth Medium (NGM).
4. Using sterile procedures, dispense 6ml of NGM into 35mm diameter petri plates using a peristaltic pump (*see* **Note 1**).
5. Leave plates at room temperature for 2 days with the lids on to allow for detection of contaminants and to allow excess moisture to evaporate. Plates can then be stored in a sealed container at 4$^o$C.

### *2.4 E. coli OP50 (food for tracking plates)*

1. Using sterile technique, streak OP50 on an LB agar plate.
2. Pick a single colony of OP50 into 50ml of LB in a 250ml Erlenmeyer flask and shake overnight at 37$^o$C.
3. Both streaked plates and liquid cultures should be stored at 4$^o$C.

## 3. Methods

### *3.1 Camera setup*

1. Run the DinoCapture software.
2. Set the camera to 640x480 resolution at 30 frames/second.
3. Navigate to the real time image settings.
4. Set the camera to grayscale (black and white "B/W" mode).
5. Turn off the camera's auxiliary white LEDs ("Aux LED Mode").
6. Save the DinoLite's configuration.
7. Quit the DinoCapture software.

8. Position the light source centered over the camera (Fig. 1B and D).
9. Position the Fresnel lens and diffuser centered over the camera (Fig. 1B and E).

### *3.2 Worm Tracker 2.0 setup*

1. Turn on the red light source and motorized stage.
2. Run the Worm Tracker 2.0 software (*see* **Note 2**).
3. On the main screen (Fig. 2C and F). Set the software to record for 5 minutes. Make sure you are logging stage coordinates and ensure that the stage position is logged when recording video.
4. Navigate to the display preferences (Fig. 3A and B). Turn on grayscale conversion and set the grayscale formula to use 100% of the red channel and 0% of the green and blue channels.
5. Navigate to the recording preferences (Fig. 4A). Restore the default file name date format, for example "_yyyy_MM_dd__HH_mm_ss".
6. Navigate to the tracking preferences (Fig. 5A and C-F). Set the tracking rate to 1 frame, the delay to 330ms, and the stage type to Zaber (assuming this is the model of motorized stage you have purchased). Set the communication port correctly so that the stage responds to software commands. Set the software to move the stage using absolute coordinates. Disable the software so it does NOT check stage responses (uncheck the box). Set the software unit conversion to 20.997 steps/micron in both axes (this value is specific to the Zaber T-NA08A50; please adjust accordingly for other stage models). Set the stage speed and acceleration to the highest values that do not cause stalling or vibration; alternatively, you may try setting the speed to 6000 (Zaber uses an arbitrary unit of measurement, roughly 0.45 microns/second) and the acceleration to 100 (Zaber uses an arbitrary unit of measurement, roughly 536 microns/second$^2$) to test whether these values work. Set the stage's home location to its center at 25,000 microns in both axes (this value is specific to the Zaber T-NA08A50; please adjust accordingly for other stage models). Set the manual stage movement size to 250 microns in both axes. Set the stage's rolling speed to 1000 microns/second in both axes. Set the software to track by centroid. Use a centroid tracking boundary of 200 microns in both axes. Set the manual threshold to 95 (unsigned 8-bit pixel intensity). Make sure the continuous auto threshold is turned off. Make sure tracking is not inverted.
7. On the main screen, save the software configuration as your default.

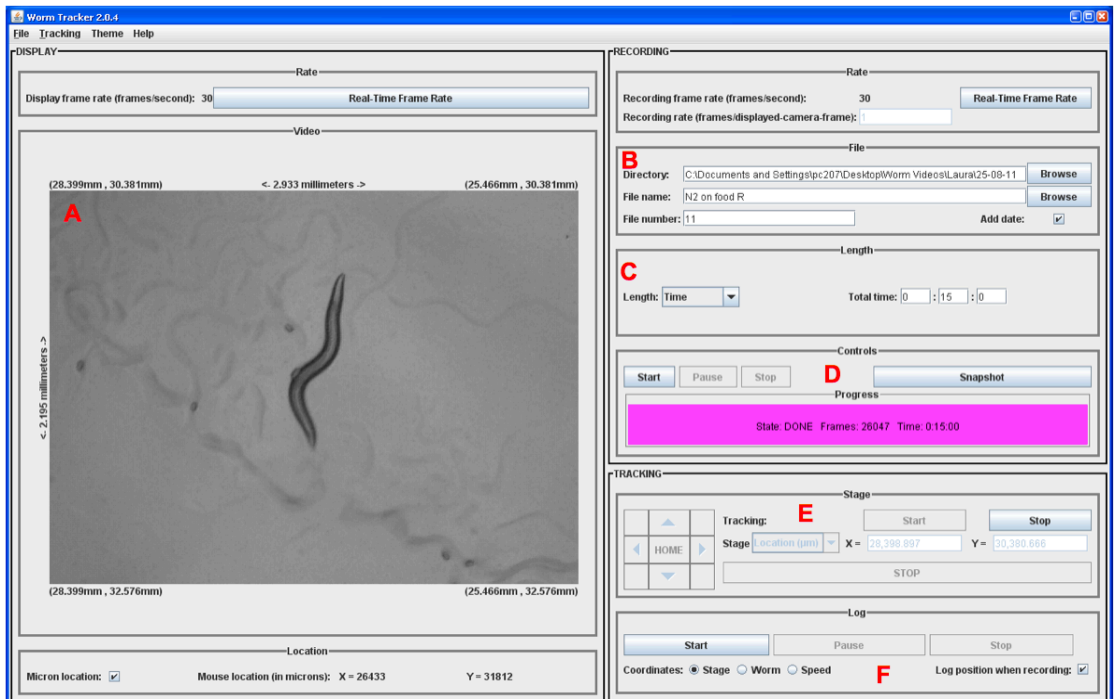

Fig. 2. WT2 Main Screen (**A**) Incoming video is displayed alongside absolute coordinates and measurements. (**B**) Experiment filename information. (**C**) Recording length information. (**D**) Recording controls. (**E**) Motorized stage and tracking controls. (**F**) Stage-movement logging controls.

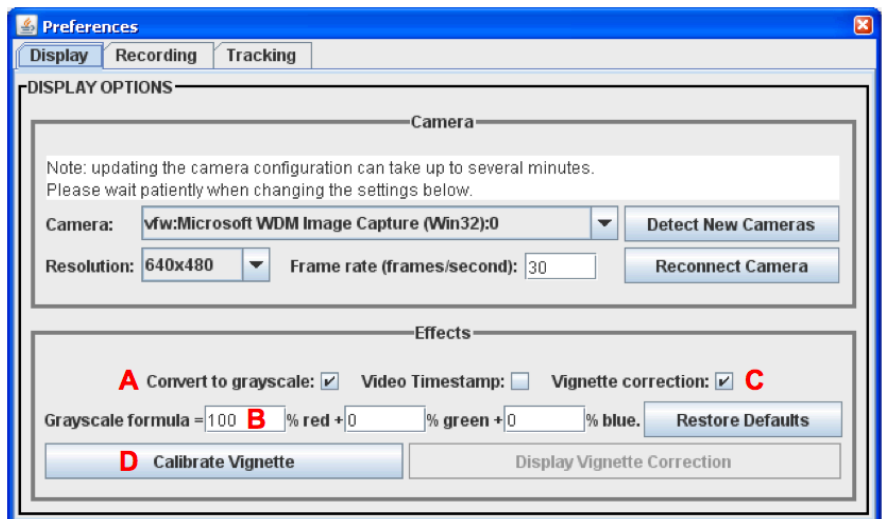

Fig. 3. WT2 Display Preferences. (A) Convert the video to grayscale. (B) The grayscale formula for converting the red, green, and blue channel in a color video to a single grayscale image. (C) Correct camera vignetting and/or dirt on the lens. (D) Calibrate the vignette correction using the current video image.

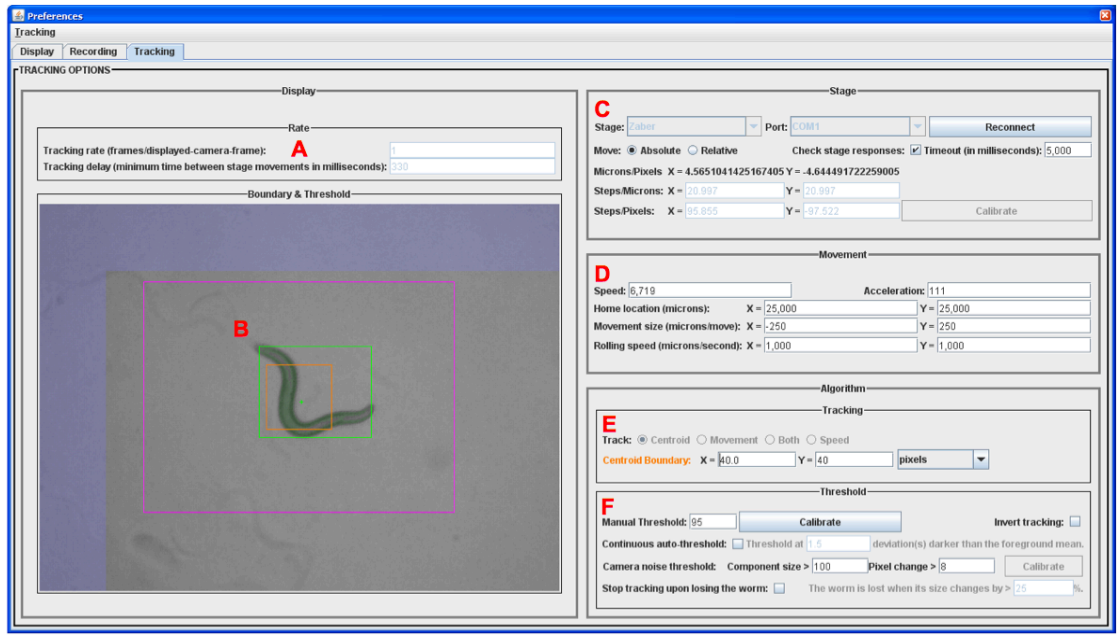

Fig. 4. WT2 Tracking Preferences. (**A**) The rate and delay for executing tracking stage movements. (**B**) The tracked worm, centroid, and MER (painted green). The centroid (painted orange) and movement (painted magenta) boundaries. Newly tracked in areas (painted lavender). (**C**) Motorized stage information. (**D**) Stage movement settings. (**E**) Tracking type (centroid and/or motion) information. F. Tracking threshold settings.

### 3.3 Camera Magnification Calibration

Ideally, the tracker should be calibrated every day before experiments are run. Alternatively, a weekly schedule of calibration can be observed with the caveat that miscalibration can lead to the loss of experimental data. Calibration should take no longer than 5-10 minutes.

1. Adjust the height of the Fresnel lens and opal diffuser so that the surface of the Fresnel lens is ~9cm from the camera lens (Fig. 1B and E).
2. Adjust the height of the light source so that the surface of the LED lens is 14mm from the camera lens (Fig. 1B and D).
3. Follow the worm and plate preparation procedures to prepare several *C. elegans* young adults on one tracking plate.
4. Place the plate face down such that the camera images the worm directly through a transparent lid. Filming through agar contributes to poor contrast and blurring.
5. Turn on the LED and stage.
6. Run the Worm Tracker 2.0 software.
7. Adjust the height of the platform (Fig. 1A) to place the plate's agar surface in focus then navigate to locate a worm (*see* **Notes 3 and 4**).
8. Adjust the camera magnification so that the worm's length is approximately ¾ the video image height (Fig. 2A).
9. Focus the worm by adjusting the worm platform height (*see* **Note 5**).
10. Adjust the height of the Fresnel lens and opal diffuser to achieve good contrast (*see* **Note 6**): the worm's contour should be dark enough to be distinguished from the plate background and food lawn (especially any worm tracks). If the contrast is too light, the worm's contour will appear to be disjoint; the contour should form a continuous black line just encasing the worm. If the contrast is too dark, the worm's contour will appear to extend well beyond its body as a surrounding shadow; furthermore, tracks in the food lawn will appear as dark as the worm's contour. Contrast is dependent on larval stage and you will need to adjust the illumination should you change the life stage of recorded worms.

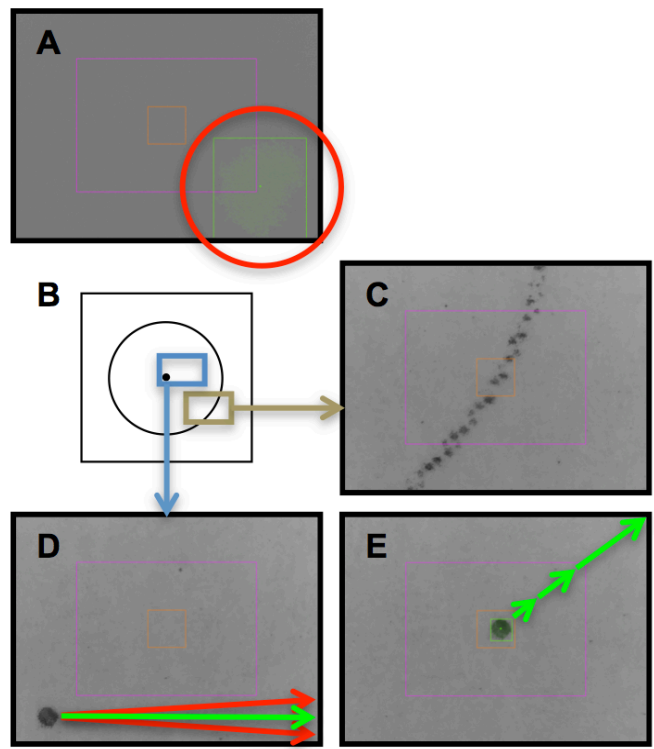

Fig. 5. WT2 Calibration Screenshots (on the Tracking Preferences screen). (A) The illumination, thresholded in green, must be centered. (B) WT2 comes with a PDF of squares, containing circles, which contain dots. A print out of the PDF is used for fine-scale calibration. (C) The circle's edge, in the print out, serves as a guide to locate the small, central dot. (D) The camera and stage must be aligned axially. X-axis stage movements must result in precise onscreen, axial dot movements (an example of axial dot movement is labeled green; the red vectors are not confined to a single axis and, therefore, indicate miscalibration). (E) WT2 auto-calibrates the scale using the known stage step size.

### *3.4 Camera-Illumination Calibration*

1. Turn on the LED and stage.
2. Swivel the Fresnel lens and opal diffuser out of the way of the light source so that the light shines directly into the camera lens. Make sure there is nothing on the worm platform.
3. Run the Worm Tracker 2.0 software.
4. Navigate to the tracking preferences.
5. Make sure automatic thresholding is turned off. Adjust the manual threshold until the camera's vignette is painted green whereas the light source remains grayscale (Fig. 5A).
6. Center the light source in the image. The vignette should be approximately equal at all four image edges. You may need to adjust the manual threshold to see the vignette throughout this procedure. Make sure the camera lens and light source are level and parallel to each other.
7. Swivel the Fresnel lens and opal diffuser back underneath the light source. Make sure the Fresnel and opal diffuser are level and parallel to the camera lens and light source.
8. Make sure you have maintained the distances between the light source, lens, and camera.
9. Make sure the camera is not focused on anything. In the Worm Tracker 2.0 display preferences, turn on the vignette correction and calibrate the vignette.
10. Save the software configuration as your default.

### *3.5 Camera-Stage Calibration*

1. Print the file "dots.pdf", located in your "Program Files\Worm Tracker\Documents" directory.
2. Cut out a square (Fig. 5B) from the print out.
3. Place the square face down onto an empty 35mm Petri plate. Then place this plate onto the worm platform. You can save this plate for future calibration.
4. Turn on the LED and stage.
5. Run the Worm Tracker 2.0 software.
6. On the main screen, locate and focus on an edge on the outer circle of your print out (Fig. 5C). If there is insufficient illumination, you may swivel the Fresnel lens and diffuser out of the way of the light source and/or turn on the camera's auxiliary white LEDs. Be sure to revert these changes when done.
7. Use the circle's outer edge to locate the central dot.
8. Use the software to move the dot. Twist the camera, in its holder, till the onscreen x and y axes match the software's manual, stage-movement commands.
9. Use the software to position the dot at one of the four onscreen corners.
10. Use the software to move the dot, solely along the x axis, to the opposite corner. If the dot changes its position in y axis, gently twist the camera, in its holder, to precisely align the camera and stage axes. Repeat this step until the software's manual, x-axis, stage-movement commands move the dot solely within the x axis (Fig. 5D).
11. Navigate to the tracking preferences.

12. Adjust the manual tracking threshold until the entire dot, and only the dot, is painted in green. Fine-tune the manual threshold so that the dot's centroid position, indicated as a green "+", remains as stable as possible.
13. Use the software to position the dot in the center of the video image (Fig. 5E).
14. Press the button to calibrate the steps/pixels and wait for the calibration to complete.
15. If the difference between the steps/pixels in both axes is more than 1, repeat steps 1-10.
16. Repeat step 14 until the steps/pixels are stable (i.e., they do not change by more than 1 from their previously calibrated values).
17. Save the software configuration as your default.
18. Navigate to the center of the motorized stages and ensure the worm platform is centered directly above the camera lens. This step ensures maximal stage coverage for the tracking plate. In other words, centering the worm platform above the camera permits the hardware to follow a worm to the edge of the plate without reaching the end of the stage's linear actuators. This can be achieved by simply navigating to the home position by pressing the "Home" button on the main screen (Fig. 2E) or using the keyboard shortcut Alt-H.

### *3.7 Worm Tracking Calibration*

1. Follow the worm and plate preparation procedure to prepare several *C. elegans* young adults on individual low-peptone NGM plates.
2. Place a plate onto the worm platform.
3. Turn on the LED and stage.
4. Run the Worm Tracker 2.0 software.
5. Locate a worm on camera by adjusting the location of the worm plate and the height of the worm platform.
6. Focus the worm by adjusting the worm platform height.
7. Navigate to the Tracking Preferences. Adjust the manual tracking threshold until the worm, and only the worm, is painted in green (Fig. 4B). Fine-tune the threshold to paint as much of the worm as possible, without bleeding into the surrounding background of worm tracks, food, and agar.
8. You may attempt to lower the tracking delay from 330ms to achieve faster tracking. But, when tracking, if you experience chaotic stage movements during which the stage continuously fails to re-center the worm, you will need to raise the delay value.
9. Save the software configuration as your default.

### *3.4 Tracking Protocol*

The tracking protocol is highly dependent upon the data being analyzed. For example, while five-minute recordings may be appropriate to measure postural properties, infrequent events such as omega turns may require much longer recording to achieve sufficient sampling. It is important to achieve good image quality for the videos. Taking effort at this stage to record high quality data will ensure that the majority of videos can be analyzed and that their extracted features are accurate. See Fig. 6 for an example of a good image contrasted with those displaying common problems.

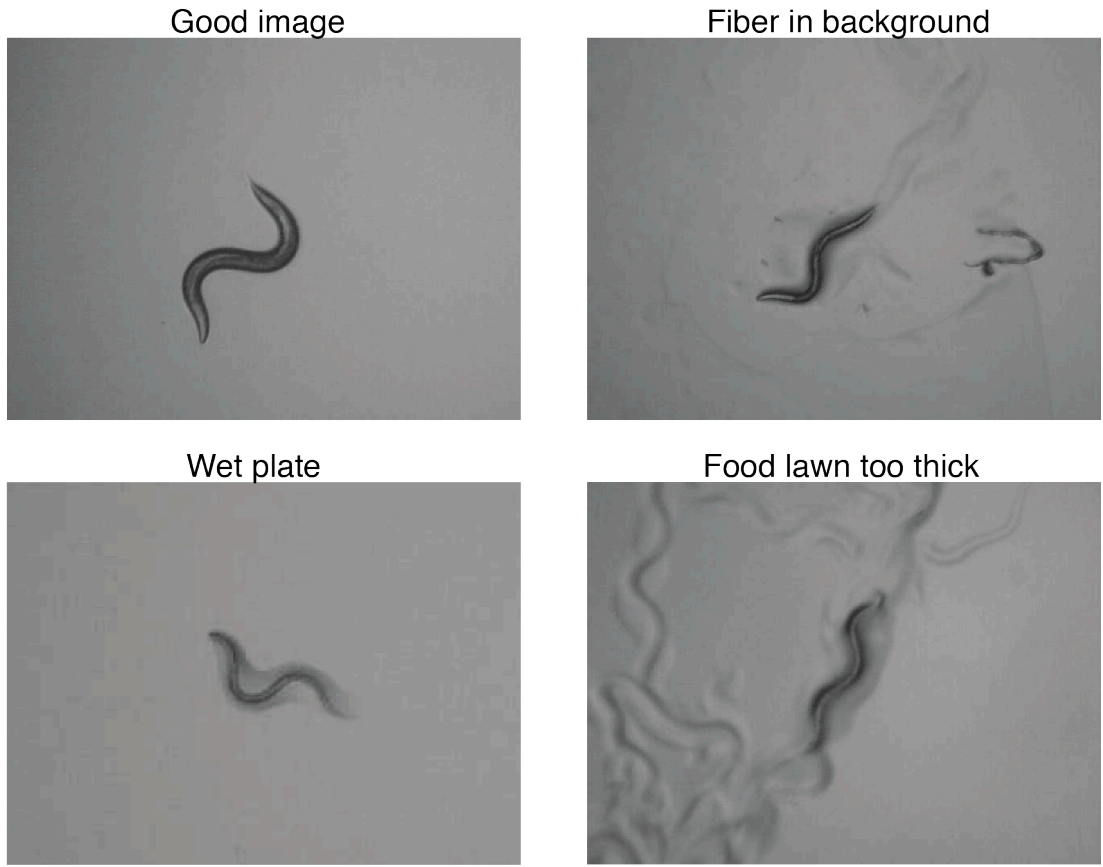

Fig. 6. Sample frames taken from a video with good contrast and magnification (top left) as well as videos taken from a plate with debris (top right), a plate that did not dry completely (bottom left), and a plate with a food lawn that is too thick (bottom right).

1. Obtain worms from the incubator. Place tracking plates and OP50 at room temperature. Discard any contaminated plates or those with imperfections such as crystals, scratches, or bubbles.
2. Resuspend the OP50 by shaking.
3. Transfer sufficient OP50 for tracking experiments to an eppendorf. To improve imaging conditions, the OP50 can be diluted by 50% for L4s to adults, 25% for L2-L3s, and 10% for L1s. Seed 20 µl onto the center of each plate. Angle the pipette when seeding plates to avoid accidentally piercing the agar. Discard any plates with imperfections in seeded bacteria.
4. Ensure that the seeded lawns on the tracking plates have dried completely, with no wet spots. At 20°C, the OP50 lawn should take ~30 minutes to dry.
5. In the worm tracking software, set the recording time.
6. Choose a tracking plate with a dry food lawn, wipe any condensation from the lid, and pick a single worm to the center of the plate taking care not to disturb the food lawn (*see* **Note 7**). Note the worm's ventral side.
7. Choose a file name and a folder to save experiments in. The following naming scheme is recommended, "<strain or worm annotation>_<room temperature>_<worm number>_<ventral side>". For example,

"N2_20C_3_AC" where "AC" indicates the ventral side is anti-clockwise from the worm's head.
8. Move the plate, agar/worm side down (facing the camera) onto the tracker and center it on the transparent imaging platform. If the plate is not centered, the stage may reach its travel limit during tracking and the worm could be lost.
9. Using the stage movement buttons (Fig. 2E) or their Alt-<arrow> keyboard shortcuts, find the worm.
10. Begin tracking by pressing the "Start" tracking button (*see* **Note 8**) or using the Alt-B keyboard shortcut and adjust the focus. Start recording (*see* **Note 9**).
11. When recording is finished, the recording "Progress" will change from yellow and pink to magenta. Each recording produces 4 associated files. These files share the same name, of the format "<user label>_<timestamp>". The 4 file suffixes are ".avi" (the video itself), ".log.csv" (a log of stage movements), ".info.xml" (XML formatted information regarding the experiment), and "vignette.dat" (a vignette correction for the video images). *See* **Note 10**.
12. When done tracking, the red LED illumination and motorized stage should be turned off and the camera lens covered for protection. Assuming, no more tracking will take place that day, the analysis should be run overnight for the worms tracked.

### 3.5 Extracting Features from Videos

1. Open Worm Analysis Toolbox.
2. Locate the "Analysis" panel and press the "Start" button.
3. Locate the "Load directory" panel and press the "Start" button.
4. Choose the directory where the worm movies are saved.
5. If you have annotated the ventral side in the filename, please check the appropriate boxes so as to inform the software.
6. Press the "Save" button to begin the analysis.
7. The analysis will run until completion and prepare feature files for each video. The feature files are stored in a "results" directory. The filename is preserved with the extension ".mat" indicating a MATLAB HDF5 formatted data file. The experiment file are stored in the hidden directory ".data" (*see* **Note 11**).
8. Quit the Worm Analysis Toolbox.

### 3.6 Converting features to summary statistics

1. Install Worm Features to Figures available at http://www.csc.mrc.ac.uk/d/Features2Figures.zip
2. Aggregate the feature files (from the "results" directory) for your control and each comparable experiment into separate folders.
3. Label the control folder "Control-<control identifier>". The control identifier is simply a string that will be used to identify it in the output.
4. Similarly, label each experiment folder "Experiment-<experiment identifier>". The program works with 1 or more experiments. Each one is compared to the same control. For multiple experiments, statistical corrections for multiple testing extend across the entire data set.

5. Run the wormFeatures2FiguresGUI program (please be patient, due to MATLAB, the program can take a long time to start up).
6. Choose the directory containing the control and experiment folders by pressing the appropriately labeled button beginning with "A".
7. Convert the feature files to figures by pressing the appropriately labeled button beginning with "B".
8. Wait for the program to complete (the window will show "***All done!" and a message will pop up). Depending on how many files you have, this can take several hours.
9. If the program gets interrupted, re-launch to start it from where it left off.
10. The "Start/End Time for Videos (in seconds)" fields allow one to run the analysis for a subset of the recorded time.
11. The program output is stored in a "tifs" folder, a "figures" folder, and a "statistics" folder. For an explanation of the tifs, see (2). The figures folder contains EPS plots with comparisons between the control and each experimental strain. In each figure, the control is darker than the experiment(s). The means are labeled as a red line. The SEMs are the blue interval around the mean. The standard deviations are the purple interval around the SEMs. The "statistics.csv" files in the statistics folder are in comma separated value format and can be opened with many programs (e.g. Excel, Numbers, R, MATLAB). Each of the 702 features is listed in rows with a corresponding number and name. After these initial columns, the control details are listed, followed by each one of the experiments. For each of these groups, summary statistics and the results of statistical tests are included. The Shapiro-Wilk test is for normality. p-values and their q-value equivalents are shown. See (2) for details of the included tests and the correction for multiple comparisons (*see* **Note 12**).

## 4. Notes

1. Using a pump to pour a consistent volume makes focusing faster when switching plates.
2. If the camera video is not displayed the problem may lie with the DinoCapture software; in this case the software should be re-installed. Alternatively, if Java has been updated, locate "Program Files\Worm Tracker\Prerequisites\JMF\jmf-2_1_1e-windows-i586.exe". Run it to uninstall then reinstall the Java Media Framework (JMF).
3. If the stage does not respond to software commands, check that the stage is connected to the correct computer port and that this port is selected in the WormTracker 2.0 tracking preferences. It is also possible that the stage manual-control knob is not in the neutral position. In this case, center the stage's actuator knobs in the neutral position (the LED on the knob will turn green). Then, reconnect the stage in the Worm Tracker 2.0 tracking preferences.
4. If the worm cannot be found, check that the agar surface is in focus. If there is insufficient range to focus on the surface, turning the plate over may help bring the surface within range. Alternatively, the worm may simply be too far from the camera's field of view. In this case, find a worm track on screen and use the stage to follow the track in order to locate the worm. Another option is to take the plate to a stereomicroscope, locate the worm noting its approximate location, then place the plate back onto the platform with the worm near the center of the camera to aid in locating it. If the worm has crawled off the agar to the plate edge, consider using undiluted OP50 for tracking.

5. If the image is still blurry after focusing, check that the camera lens is clean, that the platform is clean, and that there are no scratches anywhere occluding a clean focus. If necessary, clean the camera lens and platform or simply replace the platform if it is scratched. Check the lid for condensation and wipe clean when present.
6. Poor contrast can result from several problems. The food may be too thick. In that case, consider using undiluted OP50. If the DinoLite's auxiliary white LEDs are on, use the DinoCapture software to turn them off then save the configuration. The illumination may be off or too weak; make sure that it is on and being driven at its maximum power, 1400mA at 2.95V.
7. To ensure similar environmental conditions for controls and experiments, interleave strains (i.e. track one control worm, then one experimental worm, then repeat).
8. If the tracker fails to follow the worm, the tracking delay may be too long. It can be shortened in the preferences. The manual threshold may also be set incorrectly. In this case, repeat the calibration in section 3.7. If the stage moves chaotically, try lengthening the tracking delay within the tracking preferences and/or re-calibrating the vignette.
9. In the event that the tracker loses the worm (e.g., due to the worm reaching the edge of the plate), recording can be stopped early by pressing the "Stop" recording button. Thereafter, the associated experiment files should be discarded.
10. If recorded videos suffer dropped video frames and/or a frame rate less than 30fps, ensure other processes are not limiting the video capture rate by quitting all applications except WT2. Additionally, ensure that the DinoLite is not using a fixed exposure. This can be checked and fixed in the DinoCapture software settings. Lastly, the grayscale conversion in WormTracker 2.0 can be processor intensive and can be disabled in display preferences.
11. Errors should be ignored until the analysis finishes. Failed videos are skipped and the analysis continues to process the next video in the queue. The most common analysis complication is the "stage_movement_detection" error. This error indicates that a problem occurred matching stage movements in video to those in the log file. Unfortunately, these videos are unsalvageable and should be replaced with more experiments. Other, less frequent errors may indicate that a critical piece of software has yet to be installed (if so, reinstall the software) or that the filenames and/or paths are too long (if so, shorten the filenames and/or the directories within which they are nested).
12. If you have an interesting, uncommon result that relates to the location of the head/tail and/or dorsal/ventral side, double-check that the head/tail and dorsal/ventral side were annotated correctly. Please check the Worm Analysis Toolbox video output. Each analyzed video should be accompanied by an annotated video with the same name and an additional suffix "_seg.avi". In these videos, the head is labeled with a green dot and the ventral side is annotated by a red dot. Please skim the video and ensure that these dots correctly label the worm. Note that if the head and tail are incorrectly labeled, the dorsal/ventral side will also end up incorrect. An easy way to determine if the head and tail are correct, as long as there are less than 25 worms, is to go to the TIF file and look at the individual path traces for the worms' "Midbody Speed". Forward movement is labeled in red and reversals are labeled in blue. Most worms rarely reverse and, when they do, the reversal is rather short. As a result, path traces should be mostly in the red tones, punctuated by a few short, blue-toned reversals. It there are significant errors, discard the offending analysis files or, re-train the head/tail classifier in the Worm Analysis Toolbox and rerun it over the videos to generate correct feature files.

## Acknowledgement
This work was supported by the Medical Research Council through grant MC-A658-5TY30 to AEXB.  EIY is supported by the NIH through grant T32 MH015174- 37.